\documentclass[longbibliography]{moriond}

\usepackage[T1]{fontenc}
\usepackage{amsmath}
\usepackage{graphicx}
\usepackage{bm}
\usepackage{epsfig}
\usepackage{graphics}
\usepackage{xspace}
\usepackage{slashed}
\usepackage{xcolor}
\usepackage{longtable}
\usepackage[small]{subfigure}
\usepackage[normalem]{ulem}
\usepackage[symbol]{footmisc}
\usepackage{xspace}
\usepackage{amsmath}
\usepackage{amssymb}
\usepackage{cleveref}
\usepackage{hyperref}

\def\be{\begin{equation}}
\def\ee{\end{equation}}
\def\bea{\begin{eqnarray}}
\def\eea{\end{eqnarray}}

\makeatletter

\newcommand{\aaa}{$A + A$}

\newcommand{\pp}{$p + p$\xspace}
\newcommand{\pbpb}{$\mathrm{Pb} + \mathrm{Pb}$\xspace}

\newcommand{\auau}{$\mathrm{Au} + \mathrm{Au}$\xspace}

\newcommand{\oo}{$\mathrm{O} + \mathrm{O}$\xspace}

\newcommand{\nene}{$\mathrm{Ne} + \mathrm{Ne}$\xspace}

\newcommand{\li}{\ensuremath{{}^6\mathrm{Li}}\xspace}
\newcommand{\lili}{\ensuremath{{}^6\mathrm{Li} + {}^6\mathrm{Li}}\xspace}
\newcommand{\btbt}{${}^{10}\mathrm{B} + {}^{10}\mathrm{B}$\xspace}

\newcommand{\pdha}{$p / d / {}^3 \mathrm{He} + A$\xspace}

\newcommand{\ppb}{\ensuremath{p + \mathrm{Pb}}\xspace}

\newcommand{\hehe}[1]{${}^{#1}\mathrm{He} + {}^{#1}\mathrm{He}$\xspace}
\newcommand{\he}[1]{\ensuremath{{}^{#1}\mathrm{He}}\xspace}

\makeatother

\newcommand{\Photo}{\vspace{0.5cm}\includegraphics[height=35mm]{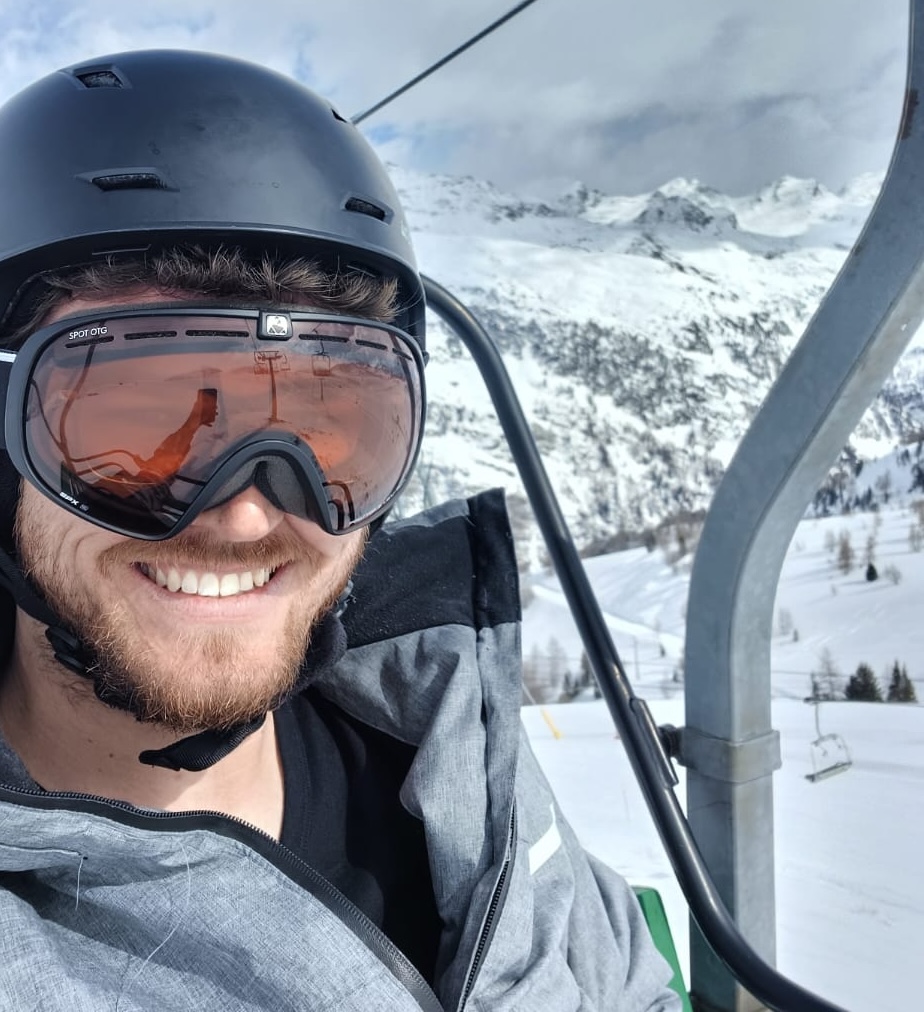}}

\begin{document}

\vspace*{4cm}

\title{Jet Quenching in the Smallest Hadronic Collision Systems\footnote{Contribution to the proceedings of the 60$^{\rm th}$ Recontres de Moriond 2026, QCD and High Energy Interactions.}}

\author{ Coleridge Faraday\footnote{Speaker, \href{mailto:frdcol002@myuct.ac.za}{frdcol002@myuct.ac.za}}$^a$, Ben Bert$^a$, Jack Brand$^a$, Werner Vogelsang$^b$, W.\ A.\ Horowitz$^{a,c,d}$ }

\address{$^a$ Department of Physics, University of Cape Town, Private Bag X3, Rondebosch 7701, South Africa\\
  \quad $^b$Institute for Theoretical Physics, University of T\"ubingen, Auf der Morgenstelle 14, D-72076 T\"ubingen, Germany\\
\quad $^c$Department of Physics, New Mexico State University, Las Cruces, New Mexico, 88003, USA\\
\quad $^d$Theoretical Sciences Visiting Program, Okinawa Institute of Science and Technology Graduate University, Onna, 904-0495, Japan
}

\maketitle\abstracts{
  We present perturbative quantum chromodynamics (pQCD) predictions for
high-momentum particle yield modification in very light ion collisions
---\btbt, \lili, \hehe{4}, and \hehe{3}---
with and without medium-induced energy loss.
We find non-trivial suppression in symmetric systems from ${}^{208}\mathrm{Pb} + {}^{208}\mathrm{Pb}$ to \hehe{3}
and in asymmetric $A+B$ systems,
with the suppression scaling approximately as $R_{AB} \simeq (\sqrt{AB})^{1/3}$.
Further, we find that \he{3} and \li{} offer particularly clean environments for observing final-state partonic energy loss
from quark-gluon plasma (QGP) formation in extremely small systems. Finally, 
we show that energy loss models generically predict $v_2\{\mathrm{SP}\} \approx 0$ in small systems, indicating that the large measured $v_2 > 0$ in \ppb is not due to energy loss.
}

\section{Introduction}

Collisions of \auau{} at RHIC and \pbpb{} at the LHC---exhibiting large low-$p_T$ $v_2$, strangeness enhancement, quarkonium suppression, and, in particular, jet quenching---have provided unequivocal evidence for quark-gluon plasma (QGP) formation.
More recently, similar signatures have emerged in high-multiplicity \pdha{} and \pp{} collisions, with the notable exception of jet quenching \cite{Grosse-Oetringhaus:2024bwr}.
To address this \textit{small-system energy loss puzzle}, \oo{} and \nene{} collisions were conducted at the LHC in 2025, yielding strong evidence for collectivity from both the soft sector, through large low-$p_T$ $v_2$ \cite{ALICE:2025luc}, and hard sector, through jet quenching \cite{CMS:2025bta}.
In light of these important results, 
two key questions remain:
why is there no observed jet quenching in \pdha{}, and what is the smallest QGP droplet?

Regarding the first question: the canonical jet quenching observable is $R_{AB} = \langle N_{\text{coll}} \rangle^{-1} Y^{AA} / Y^{pp}$, where $N_{\text{coll}}$ is the number of binary nucleon-nucleon collisions and $Y^{AA}$ ($Y^{pp}$) is the yield in \aaa{} (\pp{}) collisions, and $R_{AB} \ll 1$ indicates significant final-state energy loss. In \ppb collisions $R_{AB} \approx 1$ signifies no significant energy loss \cite{CMS:2016xef}  while 
high-$p_T$ $v_2 > 0$ indicates significant path length dependent energy loss \cite{CMS:2025kzg}.
 Regarding the second question: a natural extension of the successful \oo{} and \nene{} programmes is to consider even lighter ions at the LHC. Since bulk observables have been measured down to \pp{} collisions \cite{Grosse-Oetringhaus:2024bwr}, the key test for QGP formation in smaller systems remains jet quenching.
This work addresses both questions.

\section{Model and results}
We present predictions for high-$p_T$ $R_{AB}$ and $v_2$ from an energy loss \cite{Faraday:2023mmx,Faraday:2024gzx,Faraday:2025pto,Faraday:2025prr,Bert:2026uxa} model that computes collisional and radiative energy loss of partons traversing a 2+1D viscous hydrodynamic medium, with event-by-event and path-by-path fluctuations, followed by hadronization. 
Importantly, the model includes small-system-size corrections to the radiative \cite{Kolbe:2015rvk,Faraday:2023mmx} and collisional \cite{Faraday:2024gzx} energy loss.
Additionally, we provide baseline calculations \cite{Faraday:2025prr} for the $R_{AA}$ using NLO pQCD computations with modifications from nuclear parton distribution functions (nPDFs).
The sole free parameter, $\alpha_s$, is fixed by fitting $R_{AA}$ in \pbpb{} and \auau{} collisions. 
The model describes $R_{AA}$ as a function of $p_T$, hadron species, $\sqrt{s_{NN}}$, and centrality remarkably well \cite{Faraday:2025pto}, giving us confidence in its extrapolation to smaller systems. Indeed, energy loss models fit on large-system data generically predict non-trivial and similar suppression in small systems \cite{Faraday:2024qtl}, motivating the predictions below.

The $v_2$ is the second Fourier coefficient of the $R_{AB}$:
\begin{equation}
  R_{AB}(p_T, \phi) \equiv R_{AB}(p_T) \left(1+2 \sum_{n=1}^{\infty} v_n^{\mathrm{hard}}\left(p_T\right) \cos \left[n\left(\phi-\psi_n^{\mathrm{hard}}\left(p_T\right)\right)\right]\right)
  \label{eqn:v2def}
\end{equation}
where $\psi_{n}^{\text{hard}}$ is chosen to maximise $v_n^{\text{hard}}$. Experiments measure $v_2$ via the scalar product method or $m$-particle cumulants (equivalent in the absence of non-flow):
\begin{equation}
  v_n\{\mathrm{SP}\}\left(p_T\right) \propto \cos \left[n\left(\psi_n^{\mathrm{hard}}\left(p_T\right)-\psi_n^{\mathrm{soft}}\right)\right] v_n^{\mathrm{hard}}\left(p_T\right),
\end{equation}
where $\psi_n^{\mathrm{soft}}$ is the soft event plane angle; the proportionality constant is omitted for brevity \cite{Bert:2026uxa}.

\begin{figure}[!b]
  \centering
  \includegraphics[width=0.47\linewidth]{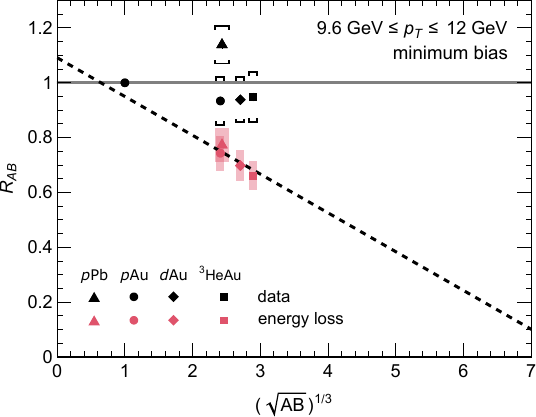}\hfill
  \includegraphics[width=0.477\linewidth]{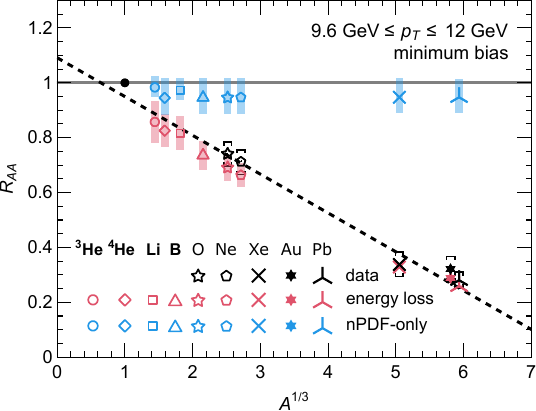}
  \caption[RAA vs system size]{Left: nuclear modification factor $R_{AB}$ as a function of $(\sqrt{AB})^{1 / 3}$ for collisions of asymmetric systems $A +B$. 
    Right: nuclear modification factor $R_{AA}$ as a function of $A^{1 / 3}$ for collisions of symmetric systems $A + A$. 
    Energy loss model predictions are shown in red, NLO pQCD nPDF-only baseline predictions shown in blue, and available data in black \cite{CMS:2016xef,PHENIX:2008saf,CMS:2018yyx,ATLAS:2022kqu,PHENIX:2021dod,CMS:2025bta,CMS:2026qef}.
    Both panels show computations for $9.6 ~\mathrm{GeV} \leq p_T \leq 12 ~\mathrm{GeV}$ $\pi^0$ hadrons produced in minimum bias collisions. 
    The same line of best fit is shown in both panels, fit to the symmetric-system data, to guide the eye. The black point at unity denotes $R_{pp} = 1$.}
  \label{fig:raa_vs_A13}
\end{figure}

\Cref{fig:raa_vs_A13} shows $R_{AB}$ vs.\ $(\sqrt{AB})^{1/3}$ for asymmetric (left) and symmetric (right) systems, with energy loss model predictions (red), NLO pQCD nPDF-only baseline (blue), and available data (black). The model describes the approximately linear $A^{1/3}$ dependence of symmetric-system $R_{AA}$ extremely well; however, extending this trend to asymmetric systems disagrees strongly with data. 
From the novel ions lighter than oxygen considered, we find that \lili{} and \hehe{3} offer unique ``goldilocks'' zones with anomalously small nPDF uncertainty and suppression.

\begin{figure}[!t]
  \centering
  \begin{minipage}{0.50\linewidth}
    \includegraphics[width=\linewidth]{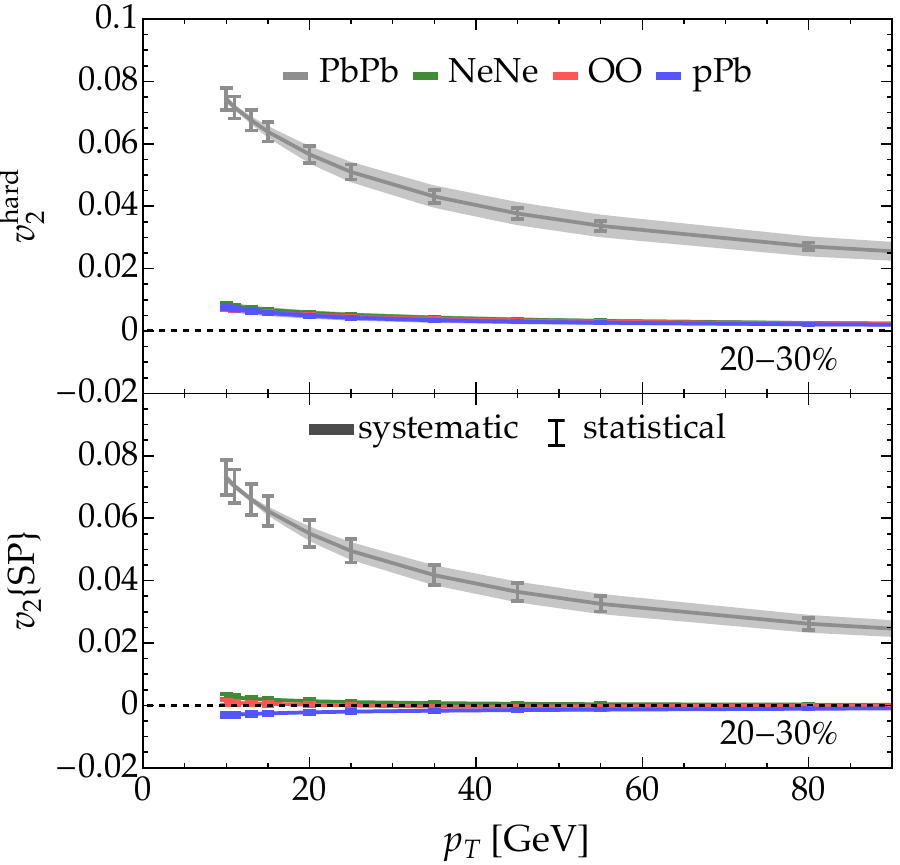}
  \end{minipage}
  \begin{minipage}{0.45\linewidth}
    \includegraphics[width=\linewidth]{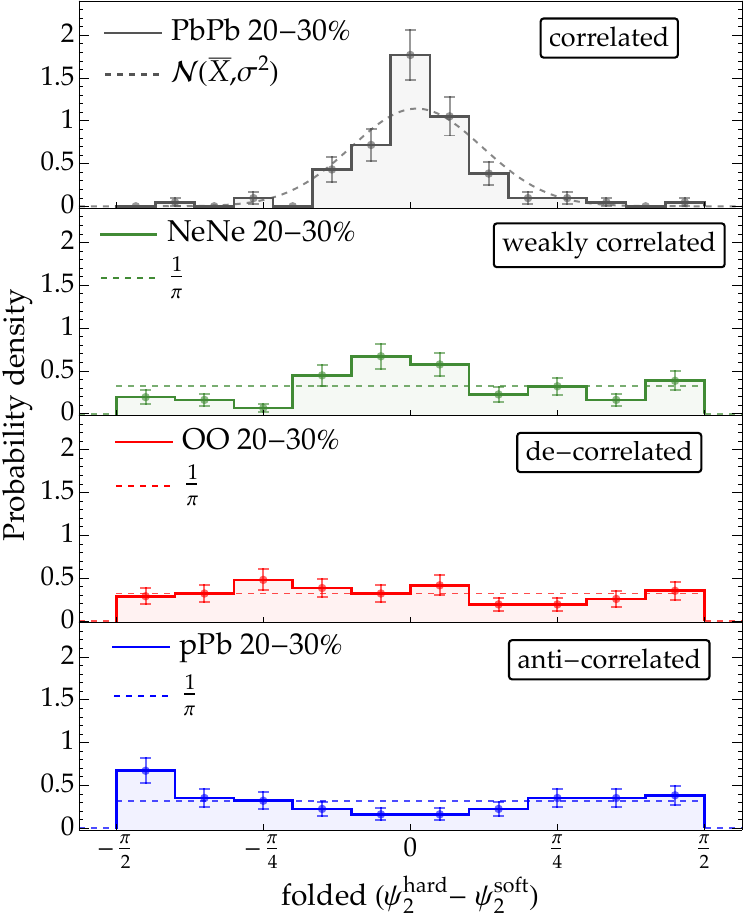}
  \end{minipage}\hfill
  \vfill
  \caption{
    Left: $v_2^{\text{hard}}$ (top) and $v_2\{\text{SP}\}$ (bottom) vs.\ $p_T$. 
    Right: 
    distribution of $\psi_2^{\text{hard}} - \psi_2^{\text{soft}}$ 
    in $20\text{--}30\%$ centrality \pbpb{} (top), \nene{} (second), \oo{} (third), and \ppb{} (bottom) collisions.
Bars denote statistical uncertainty; 
shaded bands denote systematic theoretical uncertainty.
  }
  \label{fig:v2}
\end{figure}

\Cref{fig:v2} (left) shows model predictions for $v_2^{\text{hard}}$ and $v_2\{\text{SP}\}$ vs.\ $p_T$ for \pbpb{}, \nene{}, \oo{}, and \ppb{} collisions. 
Two findings stand out: $v_2^{\text{hard}}$ is small but nonzero in all small systems, yet $v_2\{\text{SP}\}$ is nearly zero. 
\Cref{fig:v2} (right) explains this: from \cref{eqn:v2def}, $v_2\{\text{SP}\}$ depends on the correlation between hard and soft event planes. 
While \pbpb{} shows a strong correlation between $\psi_{2}^{\text{hard}}$ and $\psi_{2}^{\text{soft}}$, \nene{} and \oo{} show none, and \ppb{} shows a slight anticorrelation.
This correlation structure leads to $v_2\{\mathrm{SP}\} \approx 0$ in \nene and \oo and $v_2\{\mathrm{SP}\} \lesssim 0$ in \ppb{}.

\section{Conclusions and outlook}

We presented first predictions from our pQCD-based collisional and radiative energy loss model for $R_{AB}$ in \hehe{3}, \hehe{4}, \lili{}, and \btbt{} collisions---all candidates for future LHC runs.
We showed that $R_{AB}$ scales approximately linearly with $(\sqrt{AB})^{1 / 3}$ in our model, in agreement with available symmetric system data but in strong disagreement with available asymmetric system data, which we attribute to theoretical and experimental complications specific to asymmetric systems---for example, the non-trivial hard-soft correlations in \ppb{} collisions not included in the model \cite{Faraday:2025prr}. This discrepancy motivates the need for better controlled, symmetric small system collisions at the LHC.
Comparing energy loss suppression to nPDF-only effects, we find that \lili{} and \hehe{3} offer surprisingly clean baselines, making them ideal for searching for jet quenching in the smallest systems; a discovery of jet quenching there, in conjunction with bulk effects like $v_2$, would provide unequivocal evidence for a flowing, collective medium.

We also presented results for $v_2$ in small systems: for all systems considered---\ppb{}, \oo{}, and \nene{}---non-trivial suppression was found, yet the measurable anisotropy $v_2\{\text{SP}\}$ was nearly zero, arising from decorrelation between the hard and soft event planes, a phenomenon robust to model variations and likely generic across energy loss models.
We argue that this provides strong evidence that the large high-$p_T$ $v_2$ measured in \ppb{} is not due to energy loss but most likely due to missing physics, such as initial-state correlations, or experimental biases. Further, we provide quantitative predictions of $v_2\{\mathrm{SP}\} \approx 0$ for \oo{} and \nene{}.
Important future work includes high-$p_T$ $v_2$ measurements in \oo{} and \nene{} and an electroweak $v_2$ measurement in \ppb{}, and development of theoretical frameworks capturing both hard and soft particle production.

\section*{Acknowledgments} The authors thank Nico Orce, Giuliano Giacolone, Jamie Nagle, Petja Paakkinen, Hannu Paukkunen, John Jowett, Maciej Slupecki, Florian Jonas, Cristian Baldenegro, and Florian Lindenbauer were particularly insightful. Computations were performed using facilities provided by the University of Cape Town’s ICTS High Performance Computing team: \href{http://hpc.uct.ac.za}{hpc.uct.ac.za}. CF, BB, JB, and WAH thank the National Research Foundation and the SA-CERN collaboration for their generous financial support during the course of this work. CF thanks the National Institute for Theoretical and Computational Sciences (NITheCS) for their generous financial support.
This research was conducted in part by WAH while visiting the Okinawa Institute of Science and Technology (OIST) through the Theoretical Sciences Visiting Program (TSVP). 
WV is grateful to the Federal Ministry of Education and Research (BMFTR), grant no.\ 05P21VTCAA.

\end{document}